\documentclass[prx,twocolumn,showpacs,longbibliography,nofootinbib, superscriptaddress]{revtex4-1}

\usepackage{graphicx}
\usepackage{amssymb}
\usepackage{amsmath}
\usepackage{epstopdf}
\usepackage{color}
\usepackage[normalem]{ulem}

\newcommand{\ie}{\emph{i.e.} }

\newcommand{\eff}{\mathrm{eff}}

\newcommand{\bea}{\begin{eqnarray}}
\newcommand{\ena}{\end{eqnarray}}
\newcommand{\bee}{\begin{equation}}
\newcommand{\ene}{\end{equation}}

\newcommand{\bfig}{\begin{figure}}
\newcommand{\efig}{\end{figure}}

\newcommand{\black}{\color{black}}

\newcommand{\rxx} {\rho_{xx}}
\newcommand{\rxy} {\rho_{xy}}

\newcommand{\deln} {\Delta n}

\newcommand{\hallang} {\tan\theta_H}

\newcommand{\nbsb} {NbSb$_2$~}

\newcommand{\sxx}{S_{xx}}

\newcommand{\sxy}{S_{xy}}

\newcommand{\compfac}{\frac{\Delta n}{n_e}}

\newcommand{\hsat}{B_{\rm sat}}


\begin{document}

\title{Refining the Two-Band Model for  Highly Compensated Semimetals Using  Thermoelectric Coefficients}

\author{Ian A. Leahy$^{\dagger}$ }
\affiliation{Department of Physics, University of Colorado, Boulder, CO 80309, USA}%
\author{Andrew C. Treglia $^{\ddagger}$ }
\author{Gang Cao}
\affiliation{Department of Physics, University of Colorado, Boulder, CO 80309, USA}%
\author{Brian J. Skinner}
\affiliation{Department of Physics, Ohio State University,  Columbus,   OH  43210, USA}%
\author{Minhyea Lee}
\affiliation{Department of Physics, University of Colorado, Boulder, CO 80309, USA}%

\date{\today}
\begin{abstract}

In studying compensated semimetals,  the two-band model has proven extremely useful 
in capturing electrical conductivity under magnetic field, as a function of density and mobility of electron-like and hole-like carriers. However,  it rarely offers practical insight into magneto-thermoelectric properties.
Here, we report the field dependence of thermoelectric (TE) coefficients in a highly compensated semimetal 
NbSb$_2$, where we find the Seebeck ($\sxx$) and Nernst ($\sxy$) coefficients increase quadratically and linearly with applied magnetic field, respectively.       
Such field dependence was predicted in previous work that studied a system of two parabolic bands, within semiclassical Boltzmann transport theory when the following two conditions are simultaneously met:   
$\omega_c\tau \gg 1$ and $\tan\theta_H \ll 1$, where $\omega_c$ and $\tau$ refer to the cyclotron frequency and relaxation time, respectively, and $\theta_H$ is the Hall angle. 
 Under these conditions, we find the field dependence of the TE coefficients directly provides a relation between the electron-like ($n_e$) and hole-like ($n_h$) carrier densities, which in turn can be used to refine two-band model fitting. 
With this,  we find the compensation factor ($\frac{|\Delta n|}{n_e}$, where $\Delta n = n_e-n_h$) of NbSb$_2$ is  two orders of magnitude smaller than what was found in unrestricted fitting, resulting in a larger saturation field scale for magnetoresistance. 
Within the same framework of the semiclassical theory, we also deduce that the thermoelectric Hall angle 
$\tan\theta_{\gamma} = \frac{\sxy}{\sxx}$ can be expressed as $\big(\compfac \times \omega_c\tau\big)^{-1}$, which serves as a parameter to predict the degree of compensation.  
Our findings offer crucial insights into identifying empirical conditions for field-induced enhancement of TE performance and into engineering efficient thermoelectric devices based on semimetallic materials. 
\black
 \end{abstract}
 
\maketitle

\section{Introduction} 
\subsection{Two Band Model and Thermoelectric coefficients in a compensated semimetal}
\label{intro_sub1}
In compensated semimetals,  pockets of electron ($e$)  and hole($h$) type carriers coexist at 
the Fermi level ($E_F$). Their cancellation leads to a quadratically increasing 
magnetoresistance (MR), and an often non-linear Hall signal with an applied magnetic field ($B$)  \cite{Armitage2018, JinHu2019, Niu2021}.  Recent discoveries of enhanced thermoelectric properties in semimetals with or without topological properties of electronic structures have received a lot of attention \cite{Takahashi2016,Markov2018, Skinner2018,Markov2019, FelserReview2020}.
The two-band model (TBM) lies at the heart of describing $e-h$ compensation, offering a conceptual foundation of the band engineering by reducing the multiple pockets at $E_F$
 down to two bands, one each for $e$-like and $h$-like. 
It expresses the longitudinal ($\rxx$) and the Hall ($\rxy$) resistivities  with four intuitive parameters: carrier densities, $n_e$  for $e$-like and $n_h$ for $h$-like carriers, and corresponding mobilities ($\mu_e$ and $\mu_h$) \cite{Zimanbook_solid,AshcroftMermin} as : 
\begin{eqnarray}
\rxx&=& \frac{1}{e} \frac{(n_e\mu_e+n_h\mu_h)+(n_h\mu_e+n_e\mu_h)\mu_h\mu_eB^2}
{ (n_e\mu_e+n_h\mu_h)^2+(n_e-n_h)^2\mu_e^2\mu_h^2B^2} \label{eq:rxx} \\
\rxy&=& \frac{B}{e}\frac{(n_e\mu_e^2-n_h\mu_h^2)+(n_e-n_h)\mu_e^2\mu_h^2B^2}{(n_e\mu_e+n_h\mu_h)^2+(n_e-n_h)^2\mu_e^2\mu_h^2B^2} 
\label{eq:rxy}. 
\end{eqnarray}
In a highly compensated regime,    
the compensation factor, defined as $\frac{|\Delta n|}{n_e} = \frac{n_e- n_h}{n_e}$, is much smaller than unity, \ie $\frac{\Delta n}{n_e}\ll 1$). This results in
the fractional MR (fMR),  $\Delta\rho(B)/\rho_0 = \frac{\rxx(B)-\rho_0}{\rho_0}$, with $\rho_0 = \rxx(0)$,  increasing with $B^2$  according to Eq. (\ref{eq:rxx}). 
The saturation field scale $\hsat$, where $\Delta\rho/\rho_0$ becomes saturated, is  proportional to   $\frac{n_e}{\Delta n}$. 
Similarly, Eq. (\ref{eq:rxy}) delineates non-linear field dependences of the Hall resistivity \cite{mlee2001, Leahy2018}. 

In a typical TBM analysis,  four free parameters, 
$n_e, n_h$, $\mu_e$, and $\mu_h$ are obtained from the fit 
of the field dependences of $\rxx(B)$ and $\rxy(B)$ at a given temperature ($T$). 
They are widely utilized to characterize and to compare the electrical properties among numerous semimetals, such as  \nbsb~\cite {KWang2014, Leahy2018}, WTe$_2$~\cite{YongkangLuo2015}, 
TaAs$_2$~\cite{YongkangLuo2016},  YSb~\cite{JXu2017_YSb}, 
PtSn2~\cite{Mun2012} and  ZrP$_2$~\cite{Bannies2021}, just to name a few. 
However, determining each parameter from the four free fitting parameters without a significant degeneracy can be challenging, especially for highly compensated cases where two are close to each other, 
$n_e\approx n_h$. 
We note that $\frac{|\Delta n|}{n_e}$'s for a wide range of reported semimetals above are found to be  at least
$10^{-2}$ at low $T$'s, with one exception of  ZrP \cite{Bannies2021}.  \black
This implies the saturation field $\hsat$, where the magnetoresistance (MR) becomes field-independent, should be achieved in $10- 10^2$ T. 
Nonetheless,  the above materials have been reported not to have saturated MR. \black
This motivated us to search for a refinement of the TBM that would allow us to impose a constraint on the fitting parameters, preferably obtained from outside of the magneto-electrical conductivities measurements.

Despite its ubiquitousness in the analysis of  compensated semimetals, 
TBM has rarely been relevant to understanding the thermoelectric transport properties. 
This is simply because the Seebeck coefficient ($\sxx$) is  proportional to  
 $\compfac$, which is small  due to the $e-h$ cancellation  \cite{Zimanbook_solid}.    
In the last few years, some topological semimetals have been found to exhibit largely enhanced thermoelectric coefficients -- both Seebeck and Nernst coefficients ($\sxy$)  under applied magnetic field \cite{Rana2018, Liang2017, FelserReview2020, YPan2022, YuPan2024,FeiHan2020, WZhang2020}. 
In some of these, the enhanced field dependences are attributed to unique system-specific traits such as low  Fermi energy (hence reaching the quantum limit with a moderate strength of field), high electron mobilities, and a gapless electron spectrum that play a significant role in their TE properties as well as electrical transport  \cite{Liang2017, FeiHan2020}.

Here, we present a comprehensive analysis of the field dependence of thermoelectric (TE) coefficients of \nbsb, a textbook example of compensated semimetals with parabolic bands and high metallicity \cite{KWang2014, PengLi2022}. 
$\sxx(B)$  and $\sxy(B)$ of \nbsb~ are found to exhibit quadratic and linear dependences on the applied field, respectively, and the coefficients of $B^2$ and $B$  have strong $T$ dependence. 
This is consistent with what was predicted in the Boltzmann transport theory \cite{Feng2021} of compensated semimetals with parabolic bands. 
 In particular, the $B$-linear dependence of $\sxy$ persists even above 200 K, while the quadratic dependence of  $\sxx$  becomes hard to distinguish due to the rapid quenching of its magnitude upon increasing $T$.  
 These power-law enhancements of $\sxx(B)$ and $\sxy(B)$  are 
 {\emph {unique}} to  compensated semimetals  \cite {Feng2021}, which cannot be obtained in a single-band counterpart.  We find that the quadratic and linear coefficients with respect to $B$  play a key role in refining the TBM analysis of the electrical conductivities. 

 We will also discuss the implication on the themoelectric Hall angle $\theta_{\gamma}$, whose tangent is defined as $\tan\theta_{\gamma} =\frac{\sxy}{ \sxx-\sxx^0}$ with $\sxx^0 = \sxx(B=0)$.  
 \black

\subsection{Lorentz Drift Conditions  and their effects on thermoelectric coefficients under field}
\label{intro_sub2}
In the transverse field geometry used here, $B$ is applied perpendicular to the plane defined by the sets of current and voltage leads.
Both types of carriers are found to contribute to the heat current via Lorentz drift ($\mathbf E\times \mathbf  B$), which results in $B^2$ and $B$-linear enhancement to $\sxx(B)$ and $\sxy(B)$ based on the semiclassical calculation with the Boltzmann equation and the  Mott formula  \cite {Feng2021}.  
To facilitate the effective Lorentz drift in a compensated semimetal,  the following two conditions  \cite{Feng2021}  need to be met: 

\noindent
(i) The applied magnetic field should be large enough such that $\omega_c\tau \gg 1$, where $\omega_c= \frac{eB}{m^*}$ is the cyclotron frequency with elemental charge $e$ and the carrier effective mass $m^*$, and $\tau$ is a relaxation time.  
This condition avoids the trivial cancellation of heat current by electrons and holes.

\noindent
(ii) The Hall angle $\theta_H$ is such that $\tan\theta_H \ll 1$, where $\tan\theta_H$ is defined as $\hallang =\frac{\rxy}{\rxx}$. 
This condition implies a large positive MR  with small Hall resistivity, which will reduce the charge current at a given electric field generated by the temperature gradient.  The previous study on \nbsb \cite{Leahy2018} highlights the small  $\hallang$ under moderate strength of field along with non-saturating quadratic MR as a key attribute of  
of nearly perfectly compensated semimetals without invoking the topological nature of their electronic structures.

Achieving large $\omega_c\tau$  and small $\hallang$ simultaneously is generally 
not possible in single-band systems. However, when both conditions are met, 
the $\sxx$ and $\sxy$  of the conventional semimetals with parabolic bands grow 
with $B^2$ and $B$. 
We define two field scales $B_1$ and $B_H$ such that at $B=B_1$, $\omega\tau =1$  
and at $B=B_H$,  $\tan\theta_H = 1$. \black
In the regime of  $B_1 <B<B_H$, the field dependences of  $\sxx(B)/T$ and $\sxy(B)/T$ are expressed as 
\begin{eqnarray}
-\frac{\sxx-\sxx^0 }{T} &=&  \big(\frac{\pi g}{3}\big)^{2/3}\frac{e}{2m^*}\frac{k_B^2}{\hbar^2}\frac{\Delta n}{n_e}\frac{\tau^2}{n_e^{2/3}} B^2  \label{eq:sxx} \\
\frac{\sxy}{T} &=& \big(\frac{\pi g}{3}\big)^{2/3} \frac{k_B^2}{\hbar^2}\frac{\tau}{n_e^{2/3}}B, \label{eq:sxy}
\end{eqnarray}
where  $\sxx^0$ refers to the Seebeck coefficient at zero field (ZF), $e$  the elemental charge, $g$  the band degeneracy, and  $k_B$ the Boltzmann constant.  We assume the electrons and holes have the same transport relaxation $\tau$ and effective mass $m^*$. 

In this  study,   
 we track down the $T$ dependence of the quadratic  ($q$) coefficient 
 of the field for $\sxx(B)/T$ and 
 the linear one ($p$) for $\sxy(B)$  and compare them to Eq. (\ref{eq:sxx}) and (\ref{eq:sxy}).  
 From this, we obtain the relations among $n_e, \Delta n, \tau$ as a function of $p$ and $q$, which is used to impose a restriction on TBM fits for the longitudinal ($\rxx$) and Hall resistivities ($\rxy$). 
With these TE-bounded restrictions, our TBM fit reveals the $\compfac$ values to be of the order of  $10^{-4}$,  two orders of magnitude smaller than the standard  TBM fitting, while the two mobility values remain similar between both cases. 

 Furthermore, our refinement of TBM highlights the conditions for magnetic-field-induced enhancement of 
 thermoelectric coefficients set by $\hallang$ and $\omega_c\tau$.  
These parameters define the valid temperature and field ranges and help identify semimetal candidates with strong thermoelectric performance under magnetic field. 
\black 

\section{Experimental Methods}
\label{exp}
Single crystals of \nbsb~ were grown using the chemical vapor transport technique \cite{KWang2014}, 
resulting in high-quality single crystals of typical size of $(3-4~{\rm mm})\times(1-2~{\rm mm})\times(0.5-1$~mm). The crystal quality was confirmed with magnetotransport and XRD measurements and residual resistivity ratio \cite{Leahy2018}.

In thermoelectric coefficient measurements,  we use the configuration of the temperature gradient 
$-\nabla T \parallel  \hat x$  and an applied magnetic field $B \parallel \hat z$, where the diffusion of carriers generates an electric field $\mathbf E$ on the $xy$ plane: the Seebeck coefficient is defined as $\sxx = -E_x/|\nabla T|$ and the Nernst signal as $\sxy = E_y/|\nabla T|$.  \black
A resistive chip heater affixed to the sample provides a uniform heat current, and two pre-calibrated Cernox thermometers were used to measure  
the longitudinal $-\nabla T$.
 Voltage leads were connected to the sample 
with the Dupont silver paint with typical contact resistance $ \le  1~\Omega$.  The sample was placed in a vacuum environment  $< 10^{-5}$ mbar with a superconducting magnet to apply up to $B= 14$ T. 
The magnetic field was applied to the out-of-plane direction, while the temperature gradient and the electrical current were in the plane.


\begin{figure}
\begin{center}
\includegraphics[width= 0.8\linewidth]{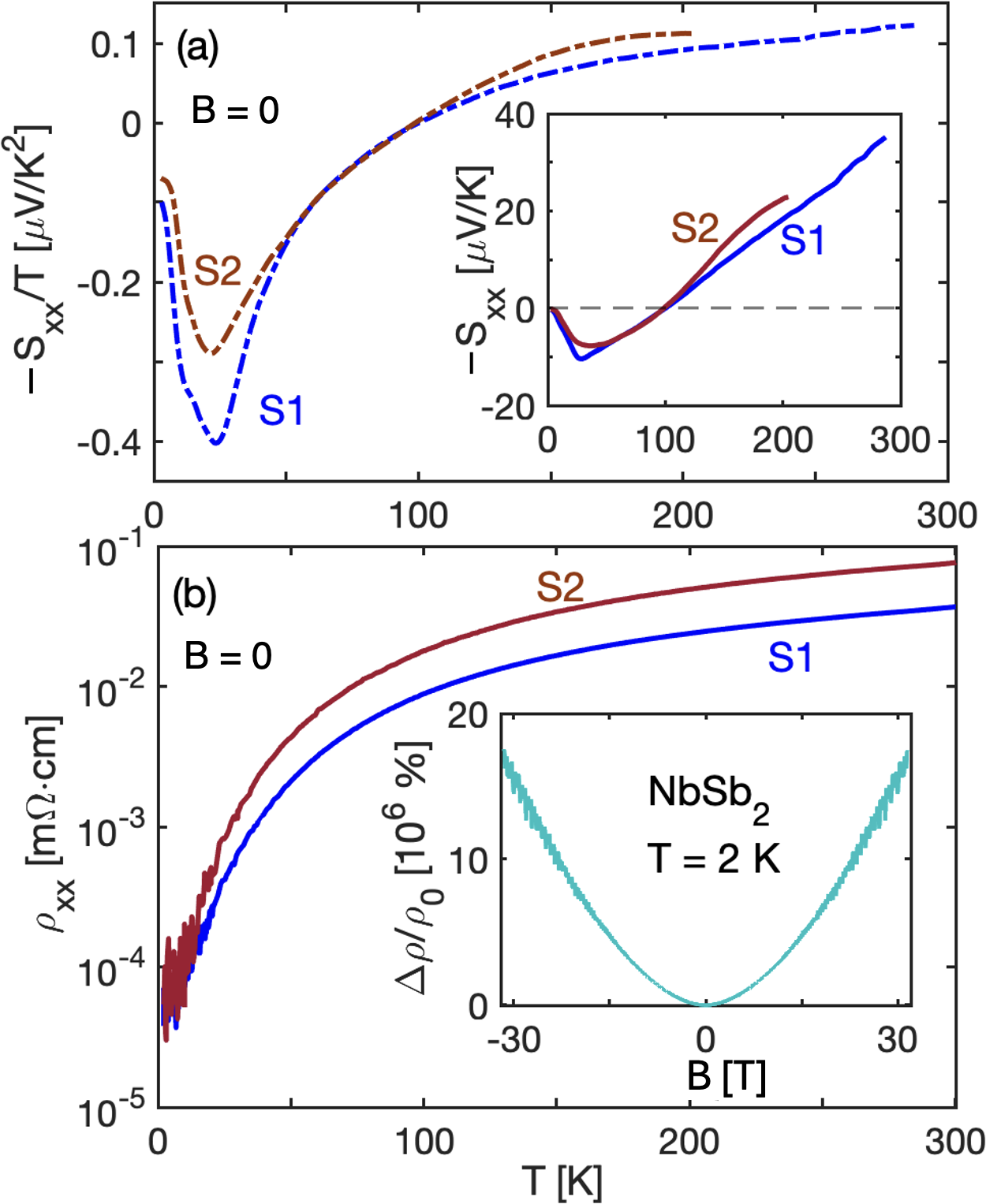}
\caption{ (a) $T$ dependence $\sxx/T$ of \nbsb for two different samples denoted as S1 and S2 at $B =0$.
$\sxx/T$ goes through zero around 100 K, and the maximum amplitude occurs at $ \approx 27$ K for both samples.
 Inset displays the $\sxx(T)$  at zero field. 
(b) Longitudinal resistivity ($\rho_{xx}$) as a function of $T$.  Inset shows the non-saturating magnetoresistance up to 31 T of \nbsb at 2 K, where the  Shubnikov-de Haas oscillation emerges at the high field in $B>15$ T. }
\label{fig:sxxrxxT}
\end{center}
\end{figure}

\section{Results}
\label{resultsec}

Fig.~\ref{fig:sxxrxxT}(a) and (b) display the temperature ($T$) dependences of 
 $-\sxx/T$ and longitudinal resistivity ($\rxx$) respectively for two \nbsb samples we used,
 denoted as S1 and S2, respectively, at zero field (ZF).    
  $-\sxx/T$ decreases monotonically until it hits zero and changes 
 its sign at $T\approx 100$ K. 
 Below 100 K, the magnitude of $\sxx/T$  reaches a maximum at $T \approx 27 $ K  and then approaches zero as $T\rightarrow$ 2 K. 
Observed magnitudes and $T$ dependences are consistent with the previous reports \cite{KWang2014, PengLi2022}.  

Panel (b) displays the resistivity ($\rxx$)  at ZF  as a function of temperature on a semilog scale.  
They exhibit a monotonic decrease as $T$ is lowered, and the decrease accelerates for $T<  50$. The residual resistivity ratio($\frac{\rho(300)}{\rho(5 {~\rm K})}$) was found to be around 700.
The inset shows the fractional magnetoresistance (fMR)  as a function of $B$ at  $T=2$ K for S1, which closely follows $B^2$  dependence according to the TBM [Eq. (\ref{eq:rxx})] \cite{Leahy2018}.  
For  $B> 17$ T,  the Shubnikov-de Haas oscillations are clearly visible.  No sign of saturation of fMR is detected up to  32 T.


\begin{figure*}[ht]
\begin{center}
\includegraphics[width= 0.85\linewidth]{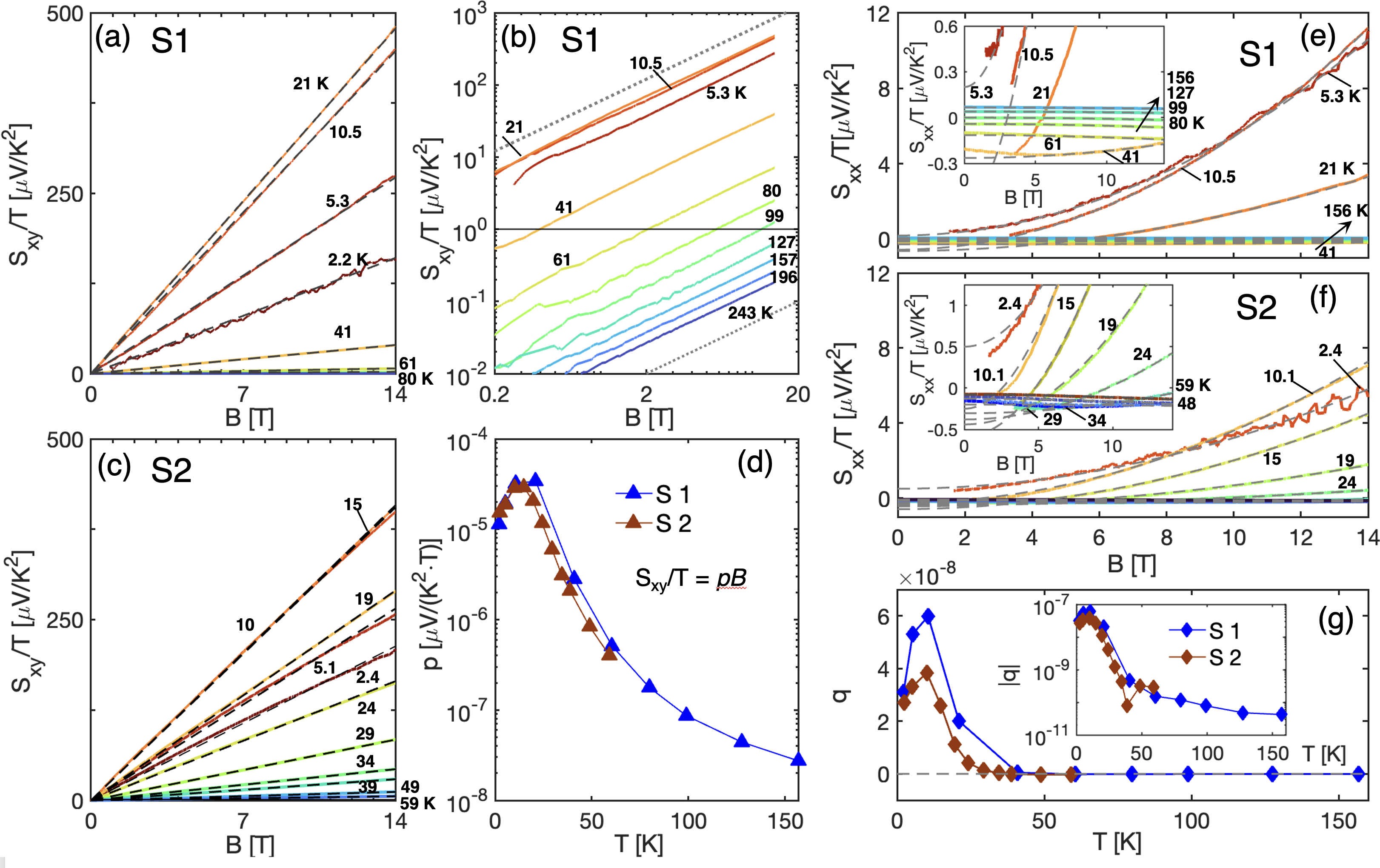}
\caption{(a)  Field dependence of  $\sxy(B)/T$ of \nbsb  for S1 are shown at various $T$'s  as indicated.  The grey broken lines indicate the linear  fit for the field dependence, where the 
(b)$\sxy(B)/T$  of S1 as a function of $B$ is shown  in $\log$-$\log$  scale.  Note that the extremely large Nernst coefficient reaches approximately $1\times10^4$ $\mu$V/K at $T=21$ K  at $B=$ 14 T, the maximum field. Dotted grey lines are guides to eyes for the linear dependence to the $B$. 
(c) $\sxy(B)/T $ are shown for S2  in $T\le60$ K. the maximum value  of $\sxy $ approximately  6000 $\mu$V/K  at $T =$ 15 K fo S2.
(d) $T$ dependence of the  linear coefficient $p(T)$  is shown in semilog scale.  
 (e-f) Seebeck coefficient  $\sxx(B)/T$ in  $B \ge 2$ T   at the same $T$s in panels (a,b)  and (c) for S1 and S2,  respectively.  Broken gray lines show the fit of the data to $-(\sxx(B)-\sxx^0) /T = q(T)B^2$, where $\sxx^0 =\sxx(B=0)$. (g) The $B$-quadratic coefficient $q(T)$ is plotted as a function of  $T$. $q(T)$, similar to $p(T)$ also quenches in $T \ge 40$ K. 
 }
\label{fig:sdata}
\end{center}
\end{figure*}


Next, we examine the field dependence of $\sxx$ and $\sxy$  for S1 and S2. Fig.  \ref{fig:sdata} displays the field dependences of the $\sxy/T$ (a-c)  and $\sxx/T$ (e-f) at various $T$'s.  
Both quantities display monotonic increases with  $B$:  For all observed $T$'s, $\sxy/T$ of both samples exhibit a linear dependence on $B$, while $\sxx/T$ exhibits a quadratic dependence. 
Thanks to the monotonic dependence of the field, the maximum value of both Seebeck and Nernst coefficient $\sxy$  occurs at the maximum field at $B=14$ T  at low temperature 
$T< 25$ K. The $T$ dependence of the maximum values of the thermoelectric coefficient will be further discussed in Fig. \ref{fig:omegactau}. 

The maximum values of  $\sxy(B)$ were found to be $1.0\times 10^4~\mu V$/K at $T=21$ K for S1,   and  $ 6.0\times 10^3~\mu$V/K at 15 K for S2.  These are consistent with previously reported values \cite{PengLi2022}.  Fig. \ref{figA1} in the Appendix show the $T$ dependence of $\sxy( B= 14$ T).  \black

The broken gray lines in Fig. \ref{fig:sdata} (a-c)  are fits to the linear relation, \ie $\sxy/T = pH$ to extract the $T$ dependence of the linear coefficient,  as shown in panel (d). 
$p(T)$ peaks at around  20 K, then plummets by two orders of magnitude by around 70 K.  

In Fig. \ref{fig:sdata}(e-f), we plot  $\sxx/T$ for S1 and S2  as a function of $B$ 
and the gray broken lines display the fit to  $-(\sxx(B)-\sxx^0)/T = q(T)B^2$, where $\sxx^0 =\sxx(B=0)$  at a given $T$ [Fig.\ref{fig:sxxrxxT} (a)] .   
Here, we fit only for $B > 2$ T  to focus on high-field behavior. The $T$ dependence of the quadratic coefficient $q$ is shown in panel (g). 
The maximum value of $\sxx(B)$ in S1  is  $1.2 \times 10^2 ~\mu$V/K at 10.5 K, while $71 ~\mu$V/K at 10.1 K in S2, both  at the max field 14 T. Fig. \ref{figA1}(b) in the Appendix  displays  the $T$ dependence of $\sxy( B= 14$ T). \black 
These values are not far from the fundamental electronic contribution $ k_B/e \approx 86~\mu$V/K. 
Similar to the case of $\sxy$, upon increasing $T$,  the magnitude of $\sxx$ begins to quench dramatically as $T$ increases above 40 K, and so does the magnitude of $q$. 
The $T$ dependence of $q(T)$ drops by three orders of magnitudes when $T$ reaches 60 K. Despite this,  we find the sign of $q$  changes from positive in low $T$  to negative at around 60 K.  With Eq. (\ref{eq:sxx}), this implies the sign of  $\Delta n$ changes.  
Insets of the panels (e) and (f) display the magnified view $\sxx(B)/T$ at elevated $T$'s as shown. 
 Although  $q(T)$ becomes very small (\ie reduced to less than   1/100 of its max)   
 and hence the $B$ dependence becomes weaker,  the $B^2$  dependences of $\sxx(B)$ 
 withhold as high as $T \ge150$ K,  except for the $T$'s where $q(T)$ crosses zero in $41 <T< 61$ K  for S1, and $34 <T< 48$ K for S2. In these temperature regions, the field dependence becomes temporarily too weak to be resolved as shown in the inset of Fig. \ref {fig:sdata} (e) and (f).  \black

The power-law field dependence of $\sxy$  and $\sxx$  in  \nbsb~  starkly 
contrast with those of other topological semimetals with smaller numbers of carriers 
with very high mobilities, such as  Cd$_3$As$_2$ \cite{Liang2017}, ZrTe$_5$ \cite{WZhang2020}, TaP \cite{FeiHan2020},  NbP \cite{CFu2018} and bismuth \cite{Spathelf2022}. In these systems,  the smaller carrier density combined with high mobility
brings down the size of  $E_F$ as well as the field scale for the extreme quantum limit  (EQL), where only the lowest Landau level is occupied at as low as  $B \sim 10^0$ T.  
Hence, the  $B$ dependence is often 
determined by the functional form of the density of states, resulting in non-monotonic and/or non-power-law behavior.   
 In \nbsb,  on the other hand,  large carrier densities and high mobilities with parabolic bands place the readily accessible magnetic field range well below the quantum limit and make the field dependence calculations straightforward \cite{Feng2021}.
 
After checking  that both TE coefficients follow  Eq. (\ref{eq:sxx}-\ref{eq:sxy}),  we analyze  $p(T)$   and $q(T)$ to connect this result to the electrical conductivity;
the coefficient $q(T)$  from $\sxx$   is expected to have  the $T$ dependence of  
$\compfac\times \frac{\tau^2}{ n_e^{2/3}}$, while $p(T)$  from $\sxy$  is proportional to $\frac{\tau}{ n_e^{2/3}}$ according to  Eq. (\ref{eq:sxx}) and Eq. (\ref{eq:sxy}), respectively. 
Putting this together, we obtain the temperature dependence of the relation between $\Delta n$ and $n_e$ 
to be
\bee
\Delta n/n_e^{1/3} = \mathcal C q/p^2, 
\label{eq:deln}
\ene
where  $\mathcal C$ is a constant comprised of $e$, $m$ and $k_B$.

Rewriting Eq.(\ref{eq:deln})  as  $n_h = n_e-\Delta n$  to 
 to eliminate $n_h$ at a given $T$, we can use this condition to fit simultaneously 
 $\rxx(B)$ and $\rxy(B)$ data  to the TBM as described  in Eq. (\ref{eq:rxx})  and  (\ref{eq:rxy}).  
 Table 1 summarizes the two TBM fitting results: the $\compfac$ in TE-bounded fitting is found to be on the order of $10^{-4}$ in low $T$ ($T\le 40$ K),  which is two orders of magnitude smaller than the standard TBM fit. We find that the values of the two carriers' mobilities are close. 
 The sign change of  $q$, inferred from the quadratic coefficients of $\sxx(B)$  [the insets of Fig. \ref{fig:sdata}(e-f)],  results in the sign changes of $\Delta n$, according to Eq.(\ref{eq:sxx}). However, the standard TBM fitting did not capture such a sign change of $\Delta n$.  
Nonetheless,  the linear coefficient of B of  $\sxy (B) $ in Eq. (\ref{eq:sxy}) is not affected by $\Delta n$'s sign change as seen in Eq. (\ref{eq:sxy})


\begin{figure}[ht]
\begin{center}
\includegraphics[width=0.8\linewidth]{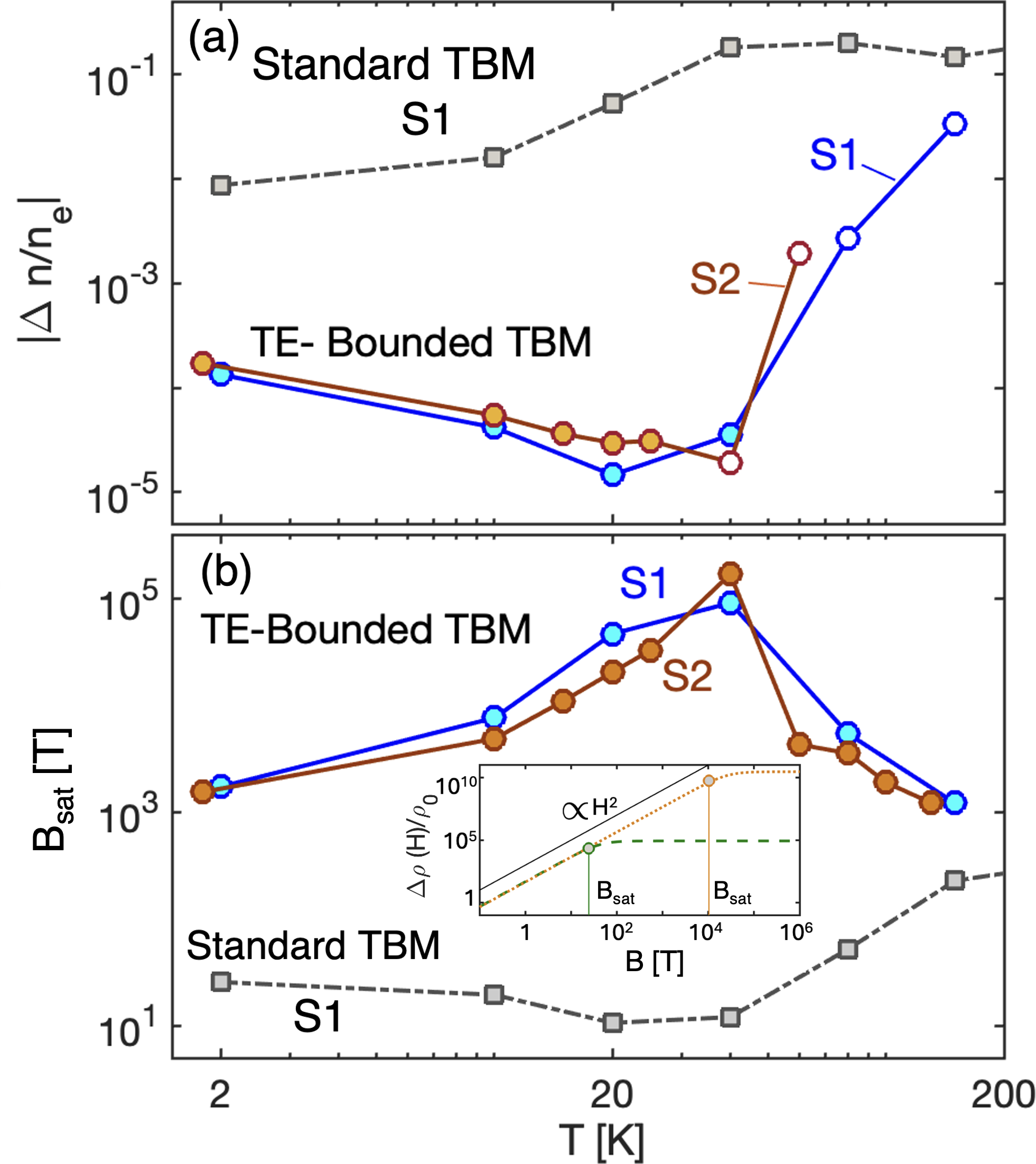}
\caption{(a) Comparison of $\deln/n_e$ obtained from the standard TBM  for S1 (square) and  $\deln/n_e$ from the TE-bounded TBM (circles)  for S1 and S2. $\compfac$  is found to be reduced by two orders of magnitude at low $T$. Open (closed) circles indicate
$\Delta n>0$ ($\Delta n< 0)$.
(b) $T$ dependence of  $\hsat$ estimated from Eq. (\ref{eq:rxx}). Inset displays the calculated fMR using values at 10 K in Table 1, with  $\compfac$'s from the standard TBM (green broken line) and from  TE-bounded (orange dotted line), where circles mark the $\hsat$, respectively.
}
\label{fig:compfac}
\end{center}
\end{figure}

 Fig. \ref{fig:compfac} (a) shows the plot of the two $\compfac$'s from Table 1 as a function of temperature. 
For $T\le40$ K,  $\Delta n/n_e$ displays a weak $T$ dependence for both standard and TE-bounded TBM with a constant discrepancy of $10^{-2}$.   Upon further increasing $T$, the pure electronic contribution becomes diluted due to strong scattering with phonons ( \ie significant reduction of $\tau$) \cite{Jaoui2020, PengLi2022}, and the difference subsides.

\begin{table*}[htb]
\begin{center}
\caption{Comparion of the two sets  of parameters of S1 obtained from the TE-bounded and the standard TBM fittings of $\rxx (B)$ and $\rxy(B)$ }
\begin{tabular}{c|cccc|cccc}
 \toprule
 & \multicolumn{4}{c} {TE-Bounded TBM} &\multicolumn {4}{c}{ Standard TBM}\\
\hline
T~[K]& $n_e[\frac{10^{26}}{\rm m^3}]$&$\frac{\Delta n}{n_e} [10^{-3}]$&$\mu_e~[T^{-1}]$&$\mu_h~[T^{-1}]$&
$n_e[\frac{10^{26}}{\rm m^3}]$&$\frac{\Delta n}{n_e}[10^{-3}]$&$\mu_e~[T^{-1}]$&$\mu_h~[T^{-1}]$\\ 
\hline
2& 4.2711 &$-0.133$&12.07&6.74&
4.2037&8.659&12.33 &6.99\\
10& 3.7114 &$-0.042$&9.35&4.72&
3.6136&15.829&9.80 &4.87\\
20& 3.1367&$-0.015$&4.85&2.09&
2.9065&52.572&5.46 &2.44\\
40& 3.1254&$-0.035$&1.05&0.44&
3.1741&181.405&1.09 &0.54\\
80& 3.0657&$+2.697$&0.24&0.09&
3.2177&199.851&0.24 &0.11\\
150& 3.1271&$+33.579$&0.09&0.03&
3.2000&146.630&0.09 &0.03\\
\hline\hline
\end{tabular}
\label{FitVals}
\end{center}
\end{table*}

A  more direct consequence of the two orders  smaller $\compfac$  
is the corresponding  larger saturation field ($\hsat$) for fMR, as it  is proportional to $\big(\frac{\Delta n}{n_e}\big)^{-1}$. 
In Fig. \ref{fig:compfac}(b),  we show the calculated   $\hsat$  using the  $\compfac$ values and the mobilities in Table 1. The inset displays the calculation of the fMR as a function of the field 
according to Eq.(\ref{eq:rxx}) for $\Delta n/n_e = 10^{-2}$ (dashed line)  
and $10^{-4}$ (dotted line)  respectively, while keeping $\mu_e$ and $\mu_h$ the same.  $\hsat$'s for each case are marked with circles.  
As displayed in Fig. \ref{fig:compfac}(b), the standard TBM calculation predicts the $\hsat$  of \nbsb to be around 20 T at 2 K, while the TE-bounded calculation predicts  $\hsat$ exceeding  $10^3$ T. As seen in the inset of Fig. 1(b), the fMR does not show any sign of saturating at its maximum field of 32 T.    
We speculate that the extreme rarity of fMR saturation observations, even at field scales exceeding predictions by the standard TBM, may stem from the imprecise estimation of $\compfac$ based on the less constrained standard fitting.  
For example,  using the standard TBM fitting, $\hsat$ for WTe$_2$ was estimated to be about  $40~\pm~2$ T  ($\compfac \approx  0.06$ \cite{YongkangLuo2015}) and   PtSn$_4$   to be $2~\pm~2$ T ($\compfac \approx  0.11$ \cite{CFu2020}), yet their fMRs display little sign of saturation far beyond these field scales,  indicating the  $\hsat$ estimated from the standard TBM is too small. 

\section{Discussion}

Our TE-constrained TBM analysis provides a crucial connection between the field-induced enhancement of the TE coefficients and the electrical conductivity. This enables prediction of  the correct $\hsat$  to understand the non-saturating MR in many compensated semimetals. Simple power-law field-dependences of $\sxx$ and $\sxy$  in the regime of $B_1<B<B_H$, as shown in Fig. \ref{fig:sdata},  is a prerequisite for this analysis. %
 As $T$  increases, the field scales of $B_1$ and $B_H$ are expected to shift, because scattering with phonons, the potential involvement of the extrinsic carriers, or other contributions that do not have an electronic origin becomes a source of $T$ dependence.  

\begin{figure}[!t]
\begin{center}
\includegraphics[width= 0.8\linewidth]{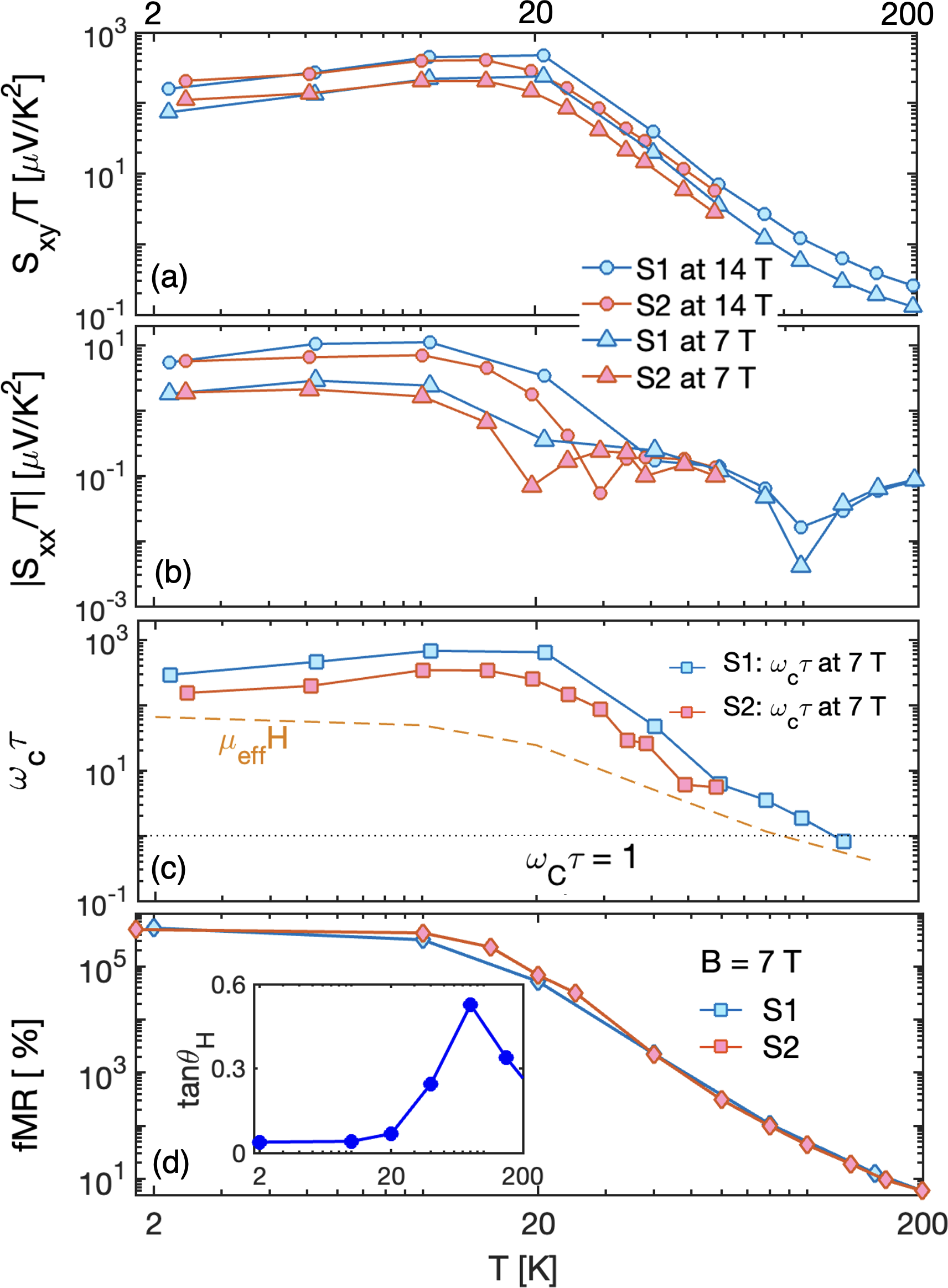}
\caption{ (a, b) $T$ dependences of $\sxy/T$ in  (a) and $\sxx/T$ in (b) at $B = 7$ T (triangles)  and 14 T  (circles) for S1 and S2. 
(c) $\omega_c\tau =\frac{eB}{m^*}\tau$ at $ B = 7$ T  is plotted as a function of $T$, using $\tau$ obtained from $p(T)$ [Fig.\ref{fig:sdata}(f) and Eq.(\ref{eq:sxy})]and the bare electron mass.  The dotted gray line  marks  $\omega_c\tau =1$. 
The orange dashed line displays $\mu_{\eff}H$ (see text) for S1. 
(d) Fractional MR ($\Delta\rho/\rho_0$)  measured at $B= 7$ T  for S1 and S2  are plotted as a function of $T$.   Similar to the $T$ dependences shown in (a-c), it decreases rapidly in $ T\ge 20$, reflecting the changes of $\tau$. 
Inset shows  $\hallang$ at $B=7$ T in the same range of $T$ for S1. illustrating the two conditions $\omega_c\tau > 1$  and $\hallang < 1$ for  Eq. (\ref{eq:sxx}) and (\ref{eq:sxy})  
are sufficiently satisfied. }
\label{fig:omegactau}
\end{center}
\end{figure}

To examine the evolution of these field scales with $T$,  we compare the $T$ dependences of various quantities in Fig. \ref{fig:omegactau}.  Panels (a) and (b) display $\sxy/T$  and $|\sxx/T|$  at $B=7$ and 14 T  as a function of $T$ in the log scale, respectively.
Both quantities remain constant at their max values in $T \le 20$ K and drop rapidly above. This is also reflected in $p(T)$ and $q(T)$ in Fig. \ref{fig:sdata}. 

The maximum value of $\sxx$ in \nbsb~ far exceeds the one expected in a metallic system with a single band,$|\sxx| \sim \frac{k_B}{e}\frac{k_BT}{E_F}$. We explain this by replacing the multiplicative factor of   $\frac{k_BT}{E_F}$ with a factor much larger than one in compensated semimetals based on Eq. (\ref{eq:sxx}), to find that the field dependence portion is offset by  both  $\omega_c\tau$ and $\compfac$
\black   
\bee
|\sxx | \propto  \frac{k_B}{e} (\omega_c\tau)^2 \compfac T, ~~~B_1\ll B \ll B_H, 
\ene 
where   $m^*$ and  $m$  refer to the effective and bare electron masses, respectively. 
With this,  the small $\compfac$ is overcome and even enhanced by 
the large $\omega_c\tau$  as shown in Fig. \ref{fig:omegactau} (c).  The relaxation time $\tau$ is obtained $\tau(T)$ from $p(T)$ using Eq. (\ref{eq:sxy}). 
We assumed  $m^*$ is the same for electron-like and hole-like carriers and is not much different from $m_e$, so that $m^*/m_e\approx 1$. 
 Hence,  $\omega_c\tau \gg 1$, one of two conditions to have enhanced TE coefficients, is sufficiently satisfied in the $T$-range where TE-coefficients exhibit the maximum values, while the other condition of $\hallang \ll 1$  has already been checked out from the magneto-electrical transport in the inset of  Fig. \ref{fig:omegactau}panel (d).

As seen in Eq.~(\ref{eq:sxx}) and (\ref{eq:sxy}),  the Nernst coefficient does not have a reduction factor of $\compfac$, but is only proportional to $ \omega_c\tau$.
This implies that  $\sxy/T$ and $\omega_c\tau$  should have the same $T$ dependence, which is consistent with our observation shown in  Fig. \ref{fig:omegactau}(a) and (c).

We also calculate  the effective mobility $\mu_{\eff}B$ at 7 T, where $\mu_{\rm eff}$  is defined as  $\mu_{\rm eff}= \frac{n_e\mu_e +n_h\mu_h}{n_e+n_h}$ using the TE-bounded values in Table 1. 
The $T$ dependence of $\mu_{\eff}B$ is displayed in Fig. \ref{fig:omegactau} (c) with an orange dashed line.  
\black
Both $\omega_c\tau$ and  $\mu_{\eff}B$ exhibit a similar  $T$ dependence, and the discrepancy in the magnitudes accounts for the different ratio of $\frac{m^*}{m_e}$  between the  $e$-like 
and the $h$-like band.

We briefly comment on the sign change of the quadratic coefficient $q(T)$;  
in the vicinity of $T \simeq 40- 60$ K,  the curvature of $\sxx(B)$  changes from  $q>0$ in low $T$ to 
$q<0$ in high $T$ [See insets of Fig. \ref{fig:sdata}(e) and (f)]. This sign change is responsible for that of $\Delta n$  as seen in Fig. 3(a).  
Combining this with the sign change of  $\sxx^0$ at ZF [Fig. \ref{fig:sxxrxxT} (a)] at around 100 K for both S1 and S2, $|\sxx/T|$  curves at 7 and 14 T [Fig.\ref{fig:omegactau} (b)]  end up having abrupt dips at different temperatures.

To connect $\omega_c\tau$ and $\compfac$ to an experimentally tangible quantity, we introduce the thermoelectric Hall angle $\theta_{\gamma}$, defined as  $\tan\theta_{\gamma} =   \frac{\sxy}{\sxx-\sxx^0}$. In compensated semimetals, it can be expressed  as 
\bee
\tan\theta_{\gamma}  \approx  (\compfac\times\omega_c\tau)^{-1}.
\ene
At $T=10$ K,  we find  $\compfac = 2.7\times 10^{-5}$ [Fig. \ref{fig:compfac}(a)] and $\omega_c\tau =  680$ [Fig. \ref{fig:omegactau} (c)]. This gives  $\tan\theta_{\gamma} = 54$.  This reasonably agrees with the calculated ratio  of  $\sxx$ ($\simeq 71~\mu$V/K for S2) and $\sxy (\simeq 4000~\mu$V/K  for S2) is found to be  57  at  $T\approx 10$ K  and $B=14 $ T. 
\black
As long as the applied magnetic field falls in the range of  $B_1<B< B_H$,  the product of $\omega_c\tau$ and $\compfac$  is a solid indicator for the overall electronic contribution to the transport quantities in compensated semimetals, and is readily estimated from the electrical conductivities alone.  Therefore, it serves as a valuable guide for identifying materials suitable for high-efficiency transverse thermoelectric applications under magnetic fields \cite{Rana2018,PengLi2022}.

As $T$ increases above 40 K, $\omega_c\tau$ decreases fast down to unity at 100 K, which indicates significant increases in scattering of charge carriers. 
This is also consistent with the rapid decrease of the fMR with $T$, shown in Fig. \ref{fig:omegactau}(d). 
The inset of panel Fig. \ref{fig:omegactau}(d) shows $\hallang$ measured at 7 T, which remains well below unity in  $T\le 20$ K.  Because  $\hallang$ of \nbsb is approximately proportional to $1/B$  in the large field limit,  the value becomes even smaller at higher fields.

All four quantities plotted in Fig. \ref{fig:omegactau} exhibit overall very similar $T$-dependence: constant at large values for the low $T$ ($T\le 20$ K), followed by rapid decreases with increasing $T$.  
The enhanced phonon scattering with charge  carriers is most likely  responsible 
for a reduction of $\tau$ resulting in the monotonic decreases  of the transport quantities, 
while we do not have evidence of any significant contribution by phonon-drag as seen in other intermetallic semiconducting compounds of  FeSb$_2$ \cite{Takahashi2016} and InAs \cite{Jaoui2020}.

We note that many semimetals reported with highly enhanced Nernst coefficients, such as NbP\cite{CFu2018}, TaP \cite{FeiHan2020}, Cd$_3$As$_2$ \cite{Liang2017}, and ZeTe$_5$ \cite{WZhang2020} are found to have  $\hallang \ge 1 $ in the moderate field range,  not to mention that their $B$ dependences are far from power-law like. 
These materials have significantly lower carrier densities than NbSb$_2$, allowing them to reach the quantum limit at lower magnetic fields. Additionally, their linear energy dispersion can give rise to field-dependent behavior distinct from what is discussed here.

The two conditions of  $\hallang \ll 1 $  and $\omega_c\tau \gg1$ can only be satisfied simultaneously in compensated semimetals, therefore serving as a key criterion in screening materials searching for enhanced TE performances under the applied field.

%


\section{Summary}
\label{sumsec}

\nbsb serves as a model system for compensated semimetals to exemplify the electronic contribution of the Nernst and Seebeck coefficients' enhancement manifested as the $B$-linear and $B$-quadratic, respectively. This is only possible when the two conditions of $\omega_c\tau \gg 1$ $\tan\theta_H \ll 1$ are simultaneously satisfied.  
Our result establishes a critical link between the field dependences of electrical and thermoelectric transport behavior,  enabling improved determination of $\compfac$ and $\hsat$  through the refined TBM. 
Our analysis also shows that the field-induced enhancement of both  TE coefficients by field in compensated semimetals can be represented by $\tan\theta_{\gamma} =\frac{\sxy}{\sxx}=\big(\compfac \times\omega_c\tau\big)^{-1}$,  a key parameter for identifying semimetallic systems with optimized field-enhanced thermoelectric performance.

\begin{acknowledgments}
 We thank  A. Fix and C. Lannert for carefully reading this manuscript. 
The syntheses of  \nbsb~ single crystalline samples used in this study and the electrical transport measurements were supported by the National Science Foundation through NSF DMR-1904361. 
The thermoelectric coefficients measurements and the data analyses by I. A. H and M. L. were supported by the  U.S. Department of Energy  Basic Energy Sciences under Award No. DE-SC0021377. 
B. J. S. was supported by the National Science Foundation under Grant No. DMR-2045742. 
Part of the electrical transport measurements were performed at the National High Magnetic Field  Laboratory, which is supported by the National Science Foundation Cooperative Agreement  DMR-1644779 and the State of Florida.


\end{acknowledgments}

\vspace{2mm}
\noindent $^{\dagger}$ Current Address: National Renewable Energy Laboratory, Golden, Colorado 80401, USA

\noindent $^{\ddagger}$ Current Address: Department of Physics, Colorado State University, Fort Collins, CO 80523, USA

\appendix*

\renewcommand{\thefigure}{A\arabic{figure}}

\setcounter{figure}{0}
\section {Temperature Dependence of the maxima of $\sxy(B=14 T)$ and $\sxx(B=14 T)$ }

\begin{figure}
\begin{center}
\includegraphics[width= .6\linewidth]{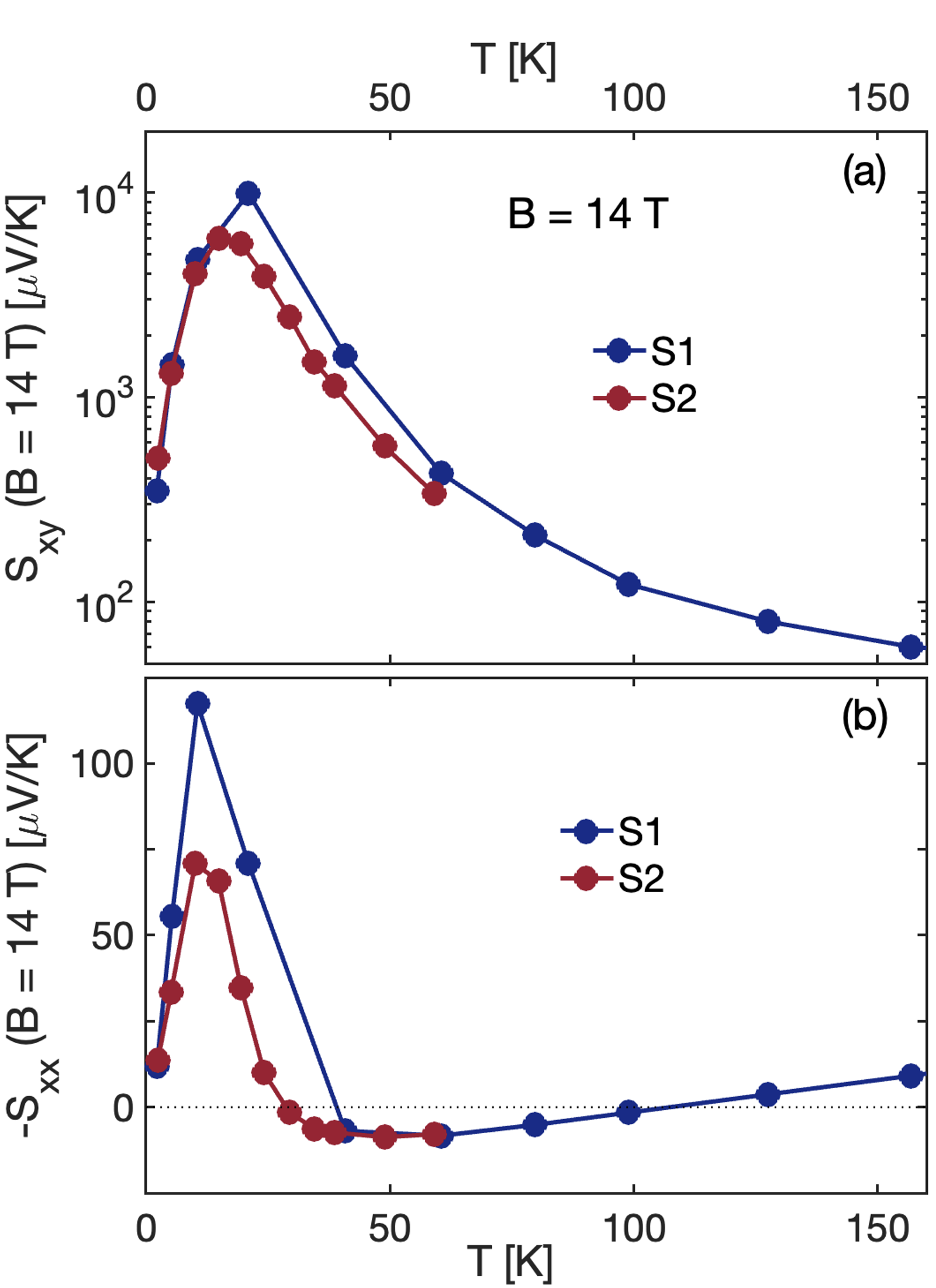}
\caption{ $T$ dependence of the  maxima values of  (a) $\sxy$ and  (b) $\sxy$  that occur at the max applied magnetic field at $B=14$ T. The dotted line in panel (a) indicates the line for $\sxx=0$. Note that the $y$-axis  is on a log scale in (a), while it is on a linear scale in (b).  }
\label{figA1}
\end{center}
\end{figure}

Fig.~\ref{figA1} (a) and (b) display the temperature dependence of the Seebeck and Nernst coefficients measured at the maximum magnetic field of 14 T.  Note that the maxima of $\sxx$ and $\sxy$ occur 
at slightly different temperatures.

%

\end{document}